\documentclass[9pt,conference]{IEEEtran}
\IEEEoverridecommandlockouts
\usepackage{amsmath,amssymb,amsfonts}
\usepackage{algorithmic}
\usepackage{textcomp}
\usepackage{xcolor}

\usepackage[]{algorithm,algorithmic}
\usepackage{cite}
\usepackage[capitalise]{cleveref}

\usepackage[pdftex]{graphicx}
\graphicspath{{../figures/}}
\DeclareGraphicsExtensions{.pdf,.jpeg,.png,.jpg}
\usepackage[export]{adjustbox}

\ifCLASSOPTIONcompsoc
 \usepackage[caption=false,font=normalsize,labelfont=sf,textfont=sf]{subfig}
\else
 \usepackage[caption=false,font=footnotesize,labelformat=empty]{subfig}
\fi

\usepackage{multirow}
\usepackage{multicol}
\usepackage{authblk}
\usepackage{balance}

\def\BibTeX{{\rm B\kern-.05em{\sc i\kern-.025em b}\kern-.08em
    T\kern-.1667em\lower.7ex\hbox{E}\kern-.125emX}}
\begin{document}
\bstctlcite{IEEEexample:BSTcontrol}
\title{LayerPipe: Accelerating Deep Neural Network Training by Intra-Layer and Inter-Layer Gradient Pipelining and Multiprocessor Scheduling}
\author{Nanda K. Unnikrishnan and Keshab K. Parhi}
\affil{Dept. Electrical and Computer Engineering, University of Minnesota\thanks{This research was supported in part by the National Science Foundation under grant number CCF-1954749.}}

\maketitle

\begin{abstract}

The time required for training the neural networks increases with size, complexity, and depth. Training model parameters by backpropagation inherently creates feedback loops. These loops hinder efficient pipelining and scheduling of the tasks within the layer and between consecutive layers. Prior approaches, such as PipeDream, have exploited the use of {\em delayed gradient} to achieve inter-layer pipelining. However, these approaches treat the entire backpropagation as a single task; this leads to an increase in computation time and processor underutilization. This paper presents novel optimization approaches where the gradient computations with respect to the weights and the activation functions are considered independently; therefore, these can be computed in parallel. This is referred to as {\em intra-layer} optimization. Additionally, the gradient computation with respect to the activation function is further divided into two parts and distributed to two consecutive layers. This leads to balanced scheduling where the computation time of each layer is the same. This is referred to as {\em inter-layer} optimization. The proposed system, referred to as {\em LayerPipe}, reduces the number of clock cycles required for training while maximizing processor utilization with minimal inter-processor communication overhead. LayerPipe achieves an average speedup of $25\%$ and upwards of $80\%$ with 7 to 9 processors with less communication overhead when compared to PipeDream.

\end{abstract}
\section{Introduction}
\label{sec:introduction}
Deep neural networks (DNNs) are omnipresent in our daily lives due to their ability to solve a wide range of complex real-world problems.
DNNs form the backbone for many tasks such as image recognition, language translation, autonomous driving, and recommendation systems~\cite{alexnet,resnet,googlenet,googleNMT17,recommender,GPT2}.
The DNN models are trained once and repeatedly used for inference. However, because of the large computations associated with training, the computation time of the training is orders of magnitude larger than that of the inference.
The massive surge in data center workloads that involve deep learning has led to new devices such as Google's TPU~\cite{TPUanalysis, TPUmotivation}, NVidia's Tesla \cite{NvidiaWP}, or Xilinx Alveo \cite{XilinxWP} in addition to custom accelerators~\cite{TNPU19,compDMA18,task_schedule18,SparseNN18,InterSCH}. 

Prior works have optimized training using parallel computations such as data parallelism~\cite{dist_survey_2019} which replicates the DNN model across processors, splits the data into multiple smaller batches, and distributes the computational workload of the model across different processors. However, data parallelization suffers from significant inter-processor communication overhead. In addition, with larger and more complex models with model parameters requiring storage in the range of GigaBytes~\cite{amoebanet}, it becomes challenging to store the model as a whole in a single processor. To overcome the issues with data parallelization, the use of {\em delayed gradients}~\cite{delayed_lms} have been exploited to achieve pipelining in~\cite{Gpipe, Pipedream} as a viable alternative. However, while these techniques effectively improve training times, they do not achieve their maximum potential due to pipeline imbalance issues and processor underutilization. This has resulted in pipeline parallelization techniques resorting back to data parallelization~\cite{Pipedream} for pipeline balancing. In this work, we propose novel {\em intra-layer} and {\em inter-layer} optimization techniques to achieve maximum processor utilization with minimal inter-processor communication overhead.

\begin{figure}[tb]
    \centering
    \includegraphics[width=\linewidth]{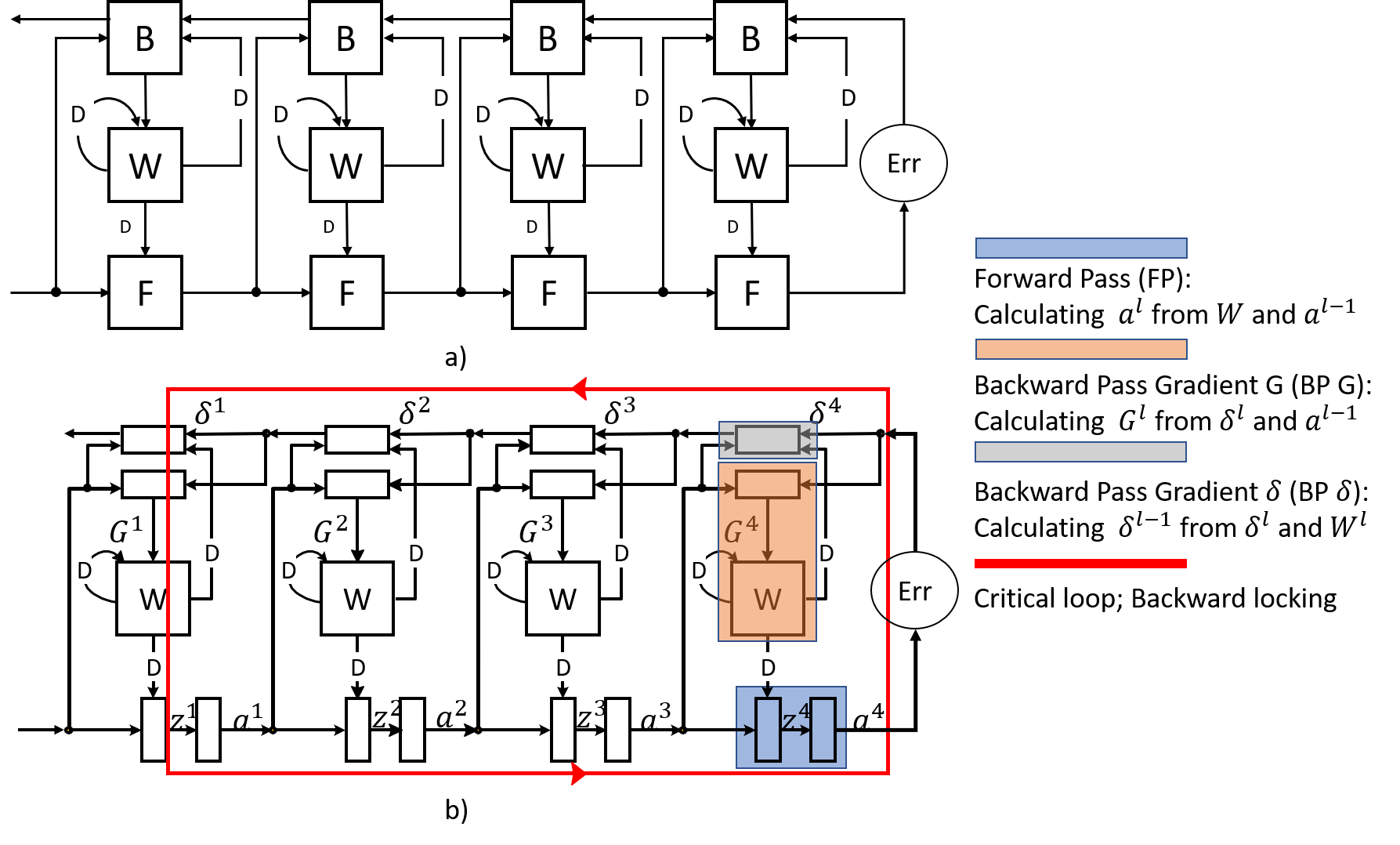}
    \caption{A sample four-layer network showing the individual operation in the forward and backward passes. a) Conventional abstraction into 3 operations: Forward, Backward and Weight update. b) Detailed view of different operations in forward and backward. The critical loop of the network is highlighted.}
    \label{fig:sample_network}
    \vspace{-1.4em}
\end{figure}

\cref{fig:sample_network}(a) shows a sample four-layer network highlighting fundamental operations of layers in the forward (F) and backward (B) passes during training. These layers can be either convolutional layers or fully-connected (FC) layers. During training, the system computes the forward pass to generate the outputs, calculates the error using ground truth, and then propagates the error backwards through the network. \cref{fig:sample_network}(b) highlights the critical loop during training which is a severe bottleneck in the computation time of DNN training. Unlike~\cite{Pipedream, Gpipe} that treat the backward computations as a single computation block as shown in \cref{fig:sample_network}(a), we split the backward pass into its fundamental operations, which involve the calculation of gradients of the error with respect to weight ($G$) and with respect to the activation function ($\delta$) as shown in \cref{fig:sample_network}(b). These two gradients can be computed in parallel using different processors. This is referred to as {\em intra-layer} parallelism. Additionally, the computation of $\delta$ is further divided into two parts which are distributed to two consecutive layers. This is referred to as {\em inter-layer} optimization.

This paper makes three key contributions listed below.
\begin{enumerate}
    \item A formal derivation of pipeline models in PipeDream~\cite{Pipedream} using concepts such as pipelining, retiming, and {\em delayed gradients}.
    \item A novel intra-layer optimization technique where the gradients of the error with respect to the weights and activation functions can be computed in parallel, reducing computation time and increasing processor utilization efficiency.
    \item A novel fine-grain inter-layer pipelining by dividing the gradient computation of the error with respect to the activation function into two parts and distributing the two parts to two consecutive layers. This leads to a balanced inter-layer pipeline and reduction of total system latency with negligible inter-processor communication overhead.
\end{enumerate}

The rest of this paper is organized as follows: Section II presents fundamental concepts of backpropagation, data, and pipeline parallelism. Section III describes the LayerPipe framework for efficient pipeline parallelism using proposed intra-layer and inter-layer optimizations. Finally, section IV evaluates and compares LayerPipe against PipeDream, a state-of-the-art pipelined accelerator architecture for training DNNs~\cite{Pipedream}.
\section{Background and prior work}
\label{sec:iter_analysis}

To optimize training time, we examine the operations within the backpropagation algorithm.

\subsection{Backpropagation}
We deconstruct the backpropagation algorithm into its primary operations to better understand how it can be implemented and optimized. The training loop for any supervised learning problem has two parts: a forward pass or \textit{inference} and a backward pass for training. \cref{fig:sample_network}(b) illustrates the data-flow graph of these operations on a sample four-layer neural network. The lower half of the data-flow graph shows the forward pass computations, while the upper half shows the backward pass computations. As shown in~\cref{fig:sample_network} there exist multiple nested feedback loops in the network. This is the main reason why system-level techniques such as pipelining are not straightforward, as delays cannot be introduced into a feedback loop system without affecting the output.
The forward and backward operations are summarized by~\cref{eq:FP,eq:FPact,eq:gradact,eq:graddelta,eq:gradW1}.
In the forward pass, the weights $\boldsymbol{W}^{l}$ and the activation output, $\boldsymbol{a}^{l-1}$, from the preceding layer are used to compute $\boldsymbol{z}^{l}$ and $\boldsymbol{a}^{l}$. The activation function $f()$ refers to non-linear activation functions such as ReLU, sigmoid or tanh. $\boldsymbol{W}^{l}$ represent the weights or filters, also referred to as model parameters, associated with the layer $l$, $\boldsymbol{z}^{l}$ is the output of the convolution operation in the convolutional layer or linear operation in the FC layer. The output $\boldsymbol{a}^{l}$ represents the activation output at layer $l$. The backward pass consists of two operations: $\boldsymbol{G}$ calculation, and $\boldsymbol{\delta}$ calculation. $\boldsymbol{G}^{l}$ is the gradient of the error function, $E$, with respect to the weights $\boldsymbol{W}^{l}$ at layer $l$, and it is computed from $\boldsymbol{\delta}^{l}$ and $\boldsymbol{a}^{l-1}$. $\boldsymbol{\delta}^{l}$ is the gradient of the error function with respect to the activations of the layer $l$ and is backpropagated to the previous layer. The notation $\odot$ represents the Hadamard product of matrices. 

\begin{equation}\label{eq:FP}
\boldsymbol{z}^{(l)} = \boldsymbol{W}^{(l)}\boldsymbol{a}^{(l-1)} 
\end{equation}
\begin{equation}\label{eq:FPact}
\boldsymbol{a}^{(l)} = f(\boldsymbol{z}^{(l)})
\end{equation}
\begin{equation}\label{eq:gradact}
\boldsymbol{\delta}^{(l-1)} = \frac{\partial E}{\partial \boldsymbol{a}^{(l-1)}} \odot f'(\boldsymbol{z}^{(l-1)})
\end{equation}
\begin{equation}\label{eq:graddelta}
where ~ \frac{\partial E}{\partial \boldsymbol{a}^{(l-1)}} = (\boldsymbol{W}^{(l) T}\boldsymbol{\delta}^{(l)})
\end{equation}
\begin{equation}\label{eq:gradW1}
{\boldsymbol{G}^{(l)}} = \left(\frac{\partial E}{\partial \boldsymbol{W}^{(l)}}\right) = \boldsymbol{\delta}^{(l)} \boldsymbol{a}^{(l-1) T}
\end{equation}

Training neural networks are characterized by the presence of several large feedback loops in the design. For example, literature has identified three fundamental {\em lockings} for the backpropagation algorithm~\cite{DNI}. First is forward locking, where a layer cannot be executed unless all the previous layers in the directed forward graph are executed. Second, update locking, a layer cannot be updated until all dependent operations have been executed in the forward pass. Last, backward locking, a layer cannot be updated before all the dependent operations are executed in forward and backward passes.
Among the three, the backward locking problem has been identified as a severe bottleneck for training and therefore has been the focus of several efforts~\cite{DNI,FR,BaPipe,FPDeep,Gpipe,Pipedream,PipeSGD}. 
\subsection{Prior work on parallelism}
Given that training times have been increasing with network size and complexity, several methods have been proposed to parallelize training.
These techniques can be broadly classified as intra-batch parallelism, and inter-batch parallelism \cite{Pipedream}.  
Intra-batch parallelism is the most common form of parallelism currently deployed to accelerate training \cite{dist_survey_2019}. In intra-batch parallelism, a single iteration of training is split across available processors.
Intra-batch techniques can be further classified into two categories: {\em data parallelism} and {\em model parallelism} \cite{dist_survey_2019, dist_survey_2020}.  The first approach, data parallelism, distributes the workload by replicating the model and splitting the input minibatch among different processors. Each model trains independently, and the gradients are updated several times before the weights.
In the second approach, model parallelism, a large model is split into multiple processors~\cite{alexnet}. Furthermore, model parallelism can split the computations along different dimensions such as channel (C), width (W), or height (H).  Lastly,  a subcategory of model parallelism is {\em layer parallelism}~\cite{DNI,FR} where the computation is split along the depth of the architecture. Model parallel approaches limit the size of the network to be stored in the processor; however, they incur large communication overheads for transferring the intermediate results, a limiting factor in systems that are communication-bounded.

Recently, pipeline parallelism has emerged as a popular technique for speeding up backpropgation~\cite{Pipedream}.
Inter-batch pipeline parallelism processes multiple mini-batches of the data in parallel over multiple iterations of the backpropagation algorithm. Intra-batch pipeline parallelism~\cite{Gpipe}, however, splits a mini-batch into smaller micro-batches to achieve the same goal.
Inter-batch pipeline parallelism is advantageous as it avoids low processor utilization by frequent synchronization. However, significant drawbacks of pipeline parallel designs include large communication overheads and difficulty balancing the workload between processors due to coarse-grained pipelining. PipeDream, for example, uses dynamic programming to find the optimal workload partition on a per-layer basis. Further works have attempted to address the workload balancing issues with fine-grain pipelining and allocation~\cite{FPDeep,BaPipe}. However, while relying on precedence graphs to derive the schedule, existing approaches do not fully exploit the precedence constraints to derive partitions that minimize communication overhead.

\begin{figure}[ht]
    \centering
    \includegraphics[width=\linewidth]{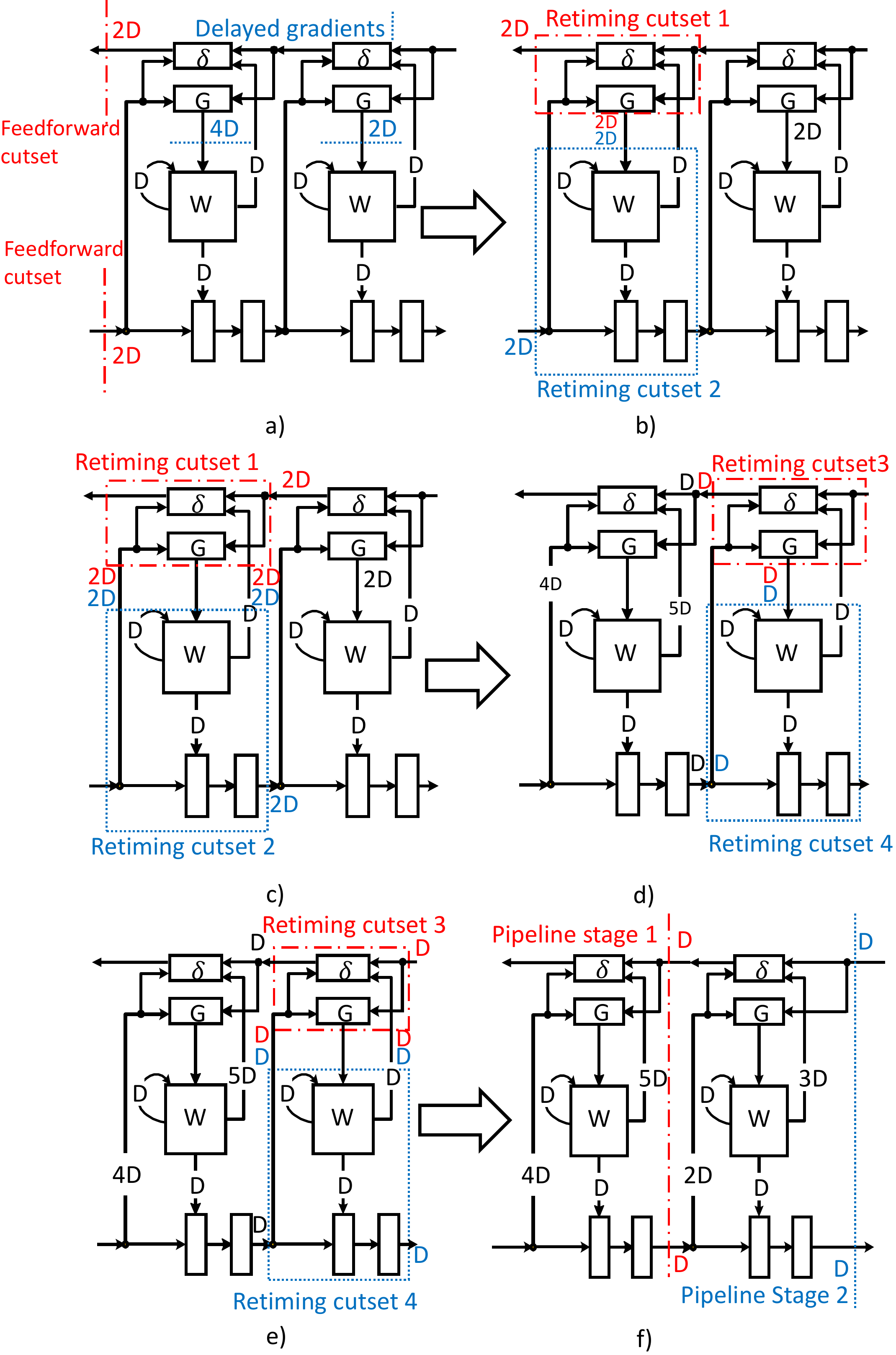}
    \caption{Step-by-step derivation (a-- f) of pipeline stages in a 2-layer design. a) Insertion of delays at feedforward cutsets and adding delayed gradients. b) Defining retiming cutsets 1 and 
    2 for the first stage. c) Pushing the delays through the retiming cutsets. d) Defining retiming cutsets 3 and 4 for stage 2. e) Pushing the delays through the retiming cutsets. f) Final pipelined DFG with two pipelined stages. }
    \label{fig:derivation}
    \vspace{-1.0em}
\end{figure}

\section{L\MakeLowercase{ayer}P\MakeLowercase{ipe}: Theory and algorithm}
\label{sec:LayerPipe}
The limitation of existing pipelined parallel algorithms in DNNs is that they do not tackle the problem from the fundamental level. This may lead to inefficient algorithms and inaccuracies as it does not account for all critical factors. However, recent improvements to tackle this problem, such as weight stashing and activation recomputation, are often rooted in rudimentary intuition and are not based on any formal theory of such systems. The following section identifies the need to formally derive these pipelined models and set up a systematic approach. 
From~\cref{fig:sample_network}, we can observe the parallels with traditional signal processing architectures, allowing us to exploit the existing well-established architecture optimizations~\cite{Parhi99}.

\subsection{Pipelining and retiming the backpropagation algorithm using delayed gradients: intra-layer optimization}

Prior work on this topic has primarily designed the system with an intuition of how the variables should behave when pipeline stages are added. For example, weight stashing~\cite{Pipedream} or activation stashing is the concept where the weights or activations are stored locally within a processor during the forward pass till the corresponding data reaches the same processor in the backward pass. In some instances, if the forward pass is computationally cheap, the activations are re-computed~\cite{Gpipe}.  Using the architecture optimization techniques from DSP systems~\cite{Parhi99}, we define two processes that must be followed to derive pipelined models: i) Locations in a system where delays can be legally inserted; ii) Necessary conditions for moving delays between edges on the data-flow graph (DFG).
For the derivation, we use a delay element to represent {\em inter-iteration dependency}~\cite{parhi_unfolding}.  Here the delay element does not refer to a physical delay but a conceptual delay. The delay holds the data for an iteration or the transition or pipeline stage that transfers data to the adjacent processor.

\begin{figure}[tb]
    \centering
    \includegraphics[width=0.95\linewidth]{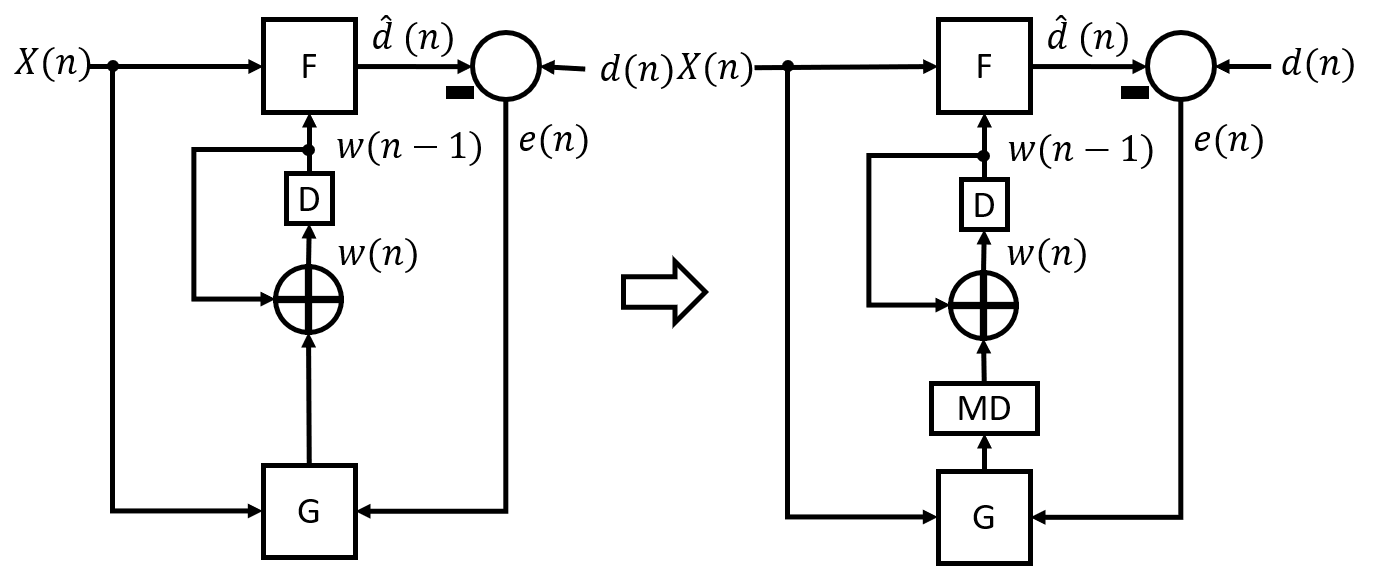}
    \caption{Delay insertion in DLMS algorithm~ \cite{delayed_lms}.}
    \label{fig:DLMS}
\end{figure}

In the first process, there are only two legal locations for delay placement within the system. Delays may be inserted through  {\em pipelining} on all feedforward cutsets; a cutset is a line that cuts along a set of edges such that it divides a graph into two subgraphs. A feedforward cutset is one where all the edges along the cutset are in the same direction. In the case of neural networks, the only feedforward cutsets are at the inputs and outputs of the network as highlighted in~\cref{fig:derivation}. Delays cannot be inserted in non-feedforward cutsets barring some exceptions. One such exception is the case of slowly varying weights in neural networks analogous to the delayed least mean squares (DLMS) algorithm~\cite{delayed_lms}. The weight update step may use an older version of the gradients in gradient descent algorithms like the DLMS algorithm. It is assumed that due to the small step size, the change in the weight with each iteration is gradual. Thus using an older or {\em stale} version of the weights or gradients will not significantly affect the convergence characteristics of the learning process. This idea forms the basis of PipeDream. We point out its equivalence with DLMS algorithm that has been widely analyzed and applied in adaptive filter applications. The weight update equation for a DLMS based approach is given by:
\begin{equation}
    \boldsymbol{W}(n) = \boldsymbol{W}(n-1) -  \eta \boldsymbol{G}(n-M)
    \label{eq:DLMS_WU}
\end{equation}
where $\boldsymbol{W}(n)$ is the weight parameter as sample $n$, $\boldsymbol{G}$ is the gradient of the error with respect to the weight parameter at sample $n$, $M$ corresponds to the degree of staleness, $\eta$ is the step size or learning rate, and $n$ is the current iteration. The $F$ block computes the forward filter and the $G$ block computes the gradient using the input $X(n)$ and the error $e(n)$. The error is the difference between the output and the ground truth (the desired value $d(n)$). This leads to the second legal location for placement of delays, between the gradient calculation and weight update states using {\em delayed gradients}. With the addition of $M$ delay elements within the feedback loop in  \cref{eq:DLMS_WU}, the system can use these elements for pipelining. As there is no synchronization step in every block, DLMS does not suffer from any overheads for the weight update operation. This process of using delayed gradients is shown in~\cref{fig:DLMS}, where MD refers to the M delays in~\cref{eq:DLMS_WU}.

The second process uses  {\em retiming}~\cite{retiming91} to move delays in the DFG to the desired location. One delay can be inserted in each outward edge of a feedback cutset during retiming if one delay is removed from each inward edge and vice-versa. Feedback cutsets are cutsets with at least one edge in both directions. At a node level, this could be the operation's outputs and inputs. Thus we can use the two processes defined above to illustrate a step-by-step process for deriving a pipelined model as shown in~\cref{fig:derivation}. Two intermediate layers of the network are chosen as a representative set for the entire network. The aim is to eventually insert a pipeline stage after the forward pass of the layer and before the backward pass of the layer.  

The first step in the process is to insert the required number of delays in the network as shown in~\cref{fig:derivation}(a). At each feedforward cutset, the total number of delays added via pipelining is the number of pipeline stages required, two in this example. At the location of delayed gradients, the number of delays to be added is twice the number of pipeline stages (or layers) after the current layer, i.e., four delays and two delays for layers 1 and 2, respectively, in this example. 

The second step uses retiming to push the required number of delays to the first pipeline stage, as shown in Figs. 2(b) and (c). We define two retiming cutsets, 1 and 2, to retime the delays highlighted in red and blue. Retiming cutset 1 pushes two delays from each outgoing edge into each incoming edge. Similarly, retiming cutset 2 pushes two delays from each incoming edge into each outgoing edge.  

Finally, in the third step, we again use retiming to move the required number of delays to the second pipeline stage, as shown in Figs. 2(d) and (e). We define two retiming cutsets, 3 and 4, used to retime the delays highlighted in red and blue. Retiming cutsets 3 and 4 are identical to cutsets 1 and 2 except that these shift one delay instead of two.  Using this, we can generalize retiming in a layer using cutsets like 1 and 2, where each cutset shifts $N$ delays, where $N$ is the number of pipeline stages after the layer.  

The final pipelined DFG is shown in~\cref{fig:derivation}(f) with its two pipelined stages and associated delay elements. The consequence of retiming was the insertion of delays between the $W$ and $\delta$ operation. Similarly, delay elements are inserted between $a$ and the backward pass calculations. These delays correspond to the weight stashing technique used by PipeDream and the activation stashing/recomputation as shown in PipeDream/Gpipe.  Thus retiming, pipelining, and delayed gradients can be used to derive any pipelined model formally. Additionally, this novel method can precisely determine how many cycles the intermediate values should be stored for multiple iterations, a result that prior works had arrived at with intuition. 

\begin{figure}[tb]
    \centering
    \includegraphics[width=\linewidth]{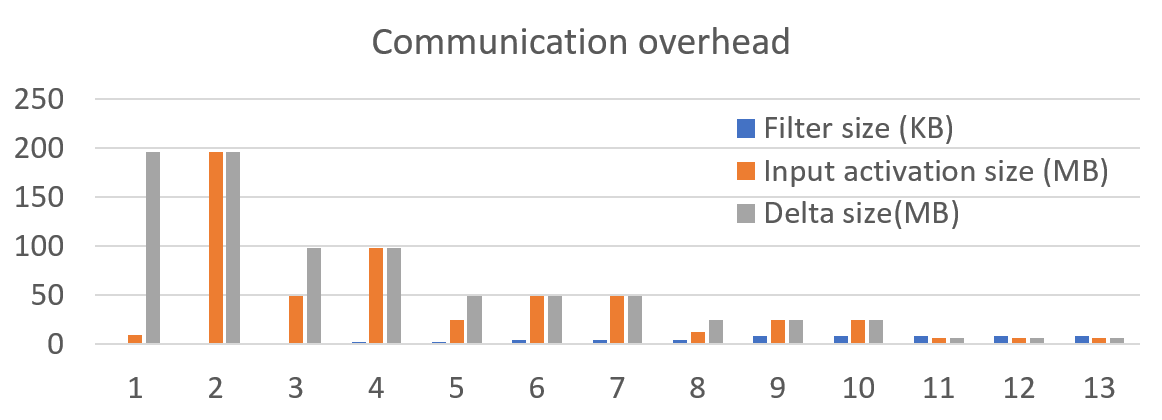}
    \caption{Layer-wise communication overhead for VGG16.}
    \label{fig:comm_overhead}
    \vspace{-1.2em}
\end{figure}

\subsection{Pipeline balancing with parallelism: inter-layer optimization}

With a formal derivation of the pipelined model, the next task is to balance the workload across multiple operations. One of the critical challenges for pipelined models is that the network architecture is not uniform, and the computation requirements across layers can vary significantly. Thus any distribution of layers across multiple processors will inevitably be imbalanced. The drawback of prior approaches is that they often lump the entire backward pass into a single operation~\cite{Pipedream,Gpipe}. Furthermore, even when fine-grain pipelining is applied~\cite{FPDeep,BaPipe}, these approaches do not exploit the characteristics of the data-flow graph (DFG) to find optimal distribution strategies.  

An examination of the data-flow graph shows that the cost to move an operation to the adjacent processor is not the same for all backpropagation operations. For example, \cref{fig:comm_overhead} shows the communication overheads associated with moving variables between different processors in the convolutional layers of VGG16.
It is observed that moving $\delta$ or $a$ between processors leads to significant communication overhead in the order of MBs. However, communicating the filters is insignificant as this transfer is in the order of KBs. In pipelined designs, at the pipeline boundaries, variables $a$ and $\delta$ have to be transferred between the processors, leading to a mandatory overhead. Any load balancing attempt will require the existing mandatory overhead, in addition to the communication required to transfer its inputs and outputs to the layers. Analyzing the backward pass in~\cref{fig:sample_network}, we can observe the input dependencies of all the operations. $\boldsymbol{G}^{l}$ computation at layer $l$ requires $\boldsymbol{a}^{l-1}$ and $\boldsymbol{\delta}^{l}$,  and $\delta$ as shown in~\cref{eq:gradW1}. Specifically~\cref{eq:graddelta} requires $\boldsymbol{\delta}^{l}$ and $W^{l}$. Note that $\boldsymbol{\delta}^{l}$ is computed in layer $l+1$. Using the example of a 2-layer network mapped to 2 processors, we explore the feasibility of shifting the computation of the backward pass of the central processor to its adjacent processors. Shifting the $G$ computation of processor $k$ to processor $k+1$ will incur significant overhead as $a_k$, activation output from processor $k$, will have to be broadcast to both processors $k$  and $k+1$. Similarly, shifting to processor $k-1$ is infeasible due to the overhead of $\delta_{k+1}$, $\delta$ output from $(k+1)$th processor. Shifting the computation of $\delta_{k}$ to processor $k+1$ will place the computation in the same processor as its input $a_{k+1}$. Also, communication of $G$ between the processors is insignificant compared to the mandatory overheads. Furthermore, the computation result is not consumed within processor $k$ and can be directly forwarded to the destination. Thus $\delta$ is a prime candidate to distribute with its adjacent processor at $k+1$. 

\begin{figure}[tb]
    \centering
    \includegraphics[width=\linewidth]{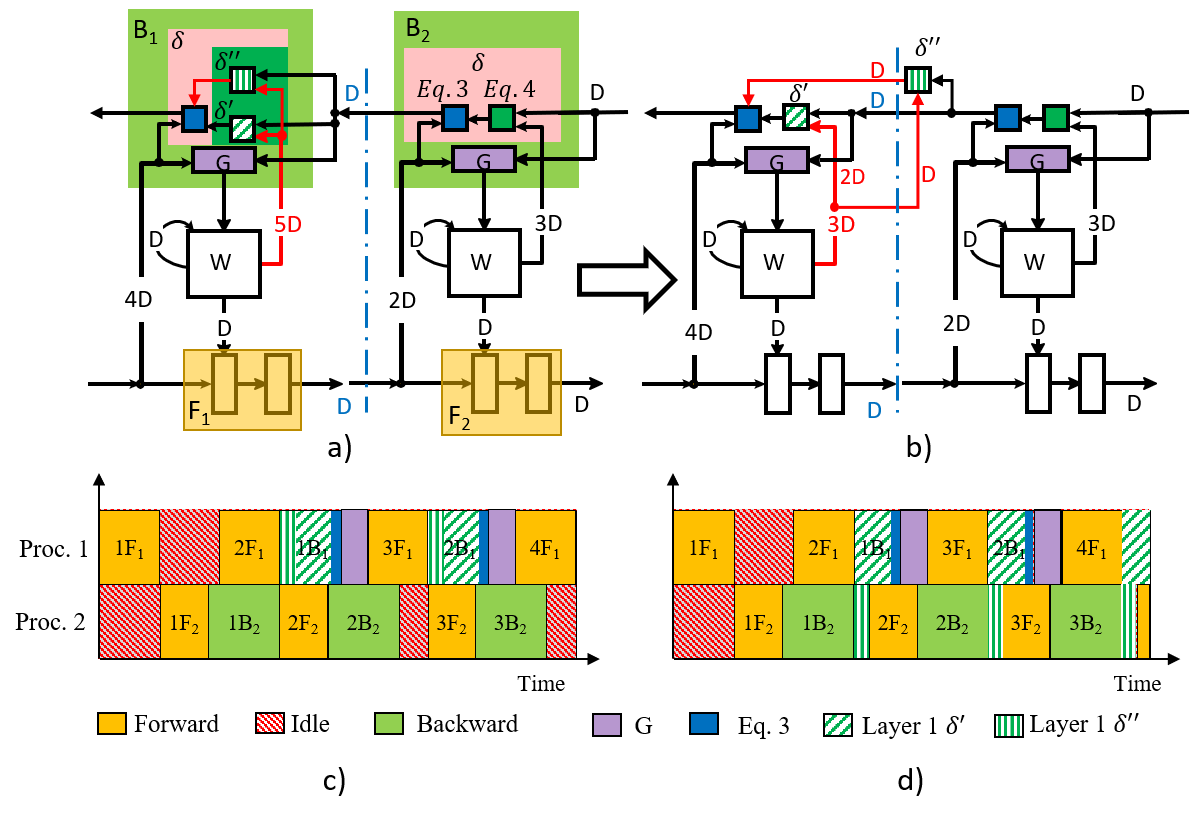}
    \caption{Block diagram of operation parallelism and its impact on scheduling. a) Original DFG with each operation mapped to its corresponding layer processor. b) Proposed DFG where $\delta''$ computation is moved to the adjacent processor. c) Imbalanced processor workloads in the original DFG. d) Balanced processor workloads due to the proposed DFG. The notation $nF_p$ indicates the forward computations for minibatch $n$ that are assigned to Processor $p$. Similarly, $nB_p$ follows the same notation for the backward computations. $nB_p$ can be further subdivided into $nB_pG$, $nB_p\delta'$, and $nB_p\delta''$.}
    \label{fig:op_parallel}
    \vspace{-1.5em}
\end{figure}

\begin{figure*}[tb]
    \centering
    \includegraphics[width=0.89\linewidth]{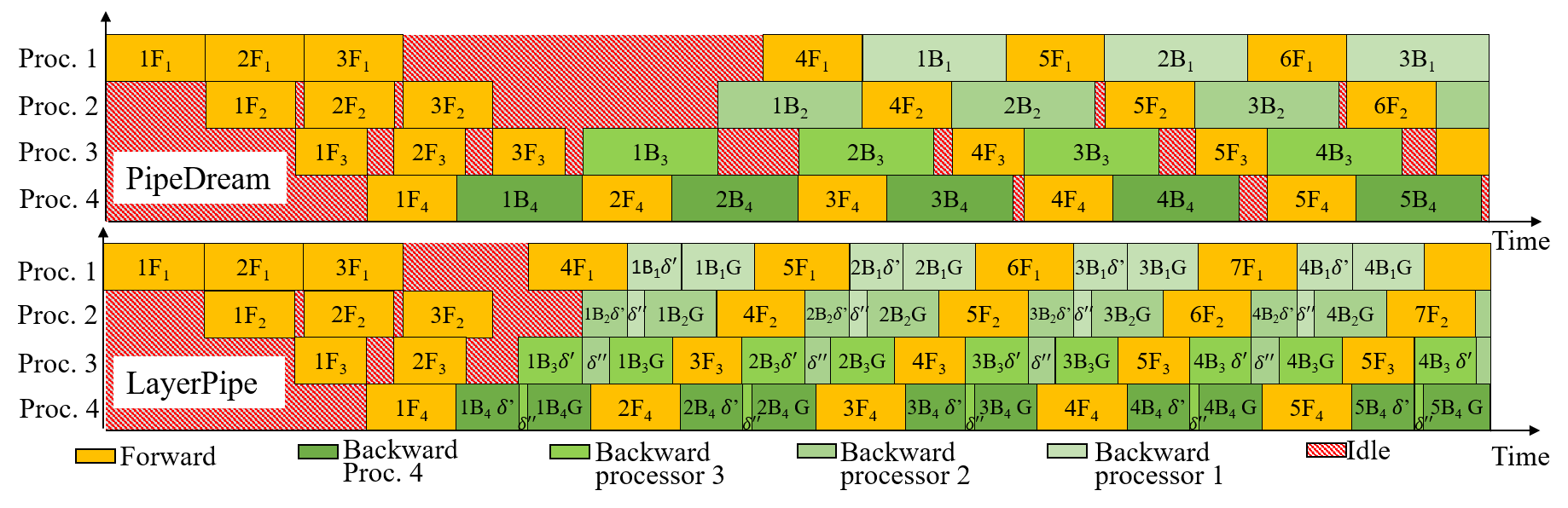}
    \caption{Contrast of scheduling approaches for PipeDream and LayerPipe.}
    \label{fig:scheduling}
\end{figure*}

\cref{fig:op_parallel} summarizes the steps required to divide $\delta$ and transfer it to the adjacent processor. In the original DFG from~\cref{fig:sample_network}, $\delta$ computations are split into 3-parts. $\delta'$ is the portion of the computation that remains within layer $l$. $\delta''$ is the portion of the computation that is shifted to layer $l+1$. In essence, layer $l+1$ {\em borrows} as much computation from layer $l$ as needed to balance the computations of consecutive processors. As $\delta$ is parallelizable in the input and output channel dimensions, it would be simple to parallelize the operation and split it between processors. This technique of dividing the $\delta$ and merging it with the adjacent layer enables {\em inter-layer} optimization. 
This approach is used to derive a load-balanced DFG as shown in~\cref{fig:derivation}(b).  In this example, the total completion time of processor 1 is much longer than processor 2, which leads to underutilization and idle time for processors. In \cref{fig:op_parallel}(b), we use retiming once again to place the delays such that $\delta'$ and $\delta''$ are computed in adjacent processors.  The current depiction highlights simple layer connections; however, there can be multiple ways these brain-inspired neural networks can be connected~\cite{ojcas}, including divergent or convergent paths. In examples such as ResNet or U-net~\cite{unet} with convergent paths for $\delta$, an additional summation step is required before the result can be used; thus, $\delta''$ would only be moved to the processor that computes this step.  

Figs. 5(c) and (d) represent the comparison between a layer-based coarse and fine-grained layer parallelism. In \cref{fig:op_parallel}(c), the difference in computation times between the processors is approximately a third of the computation time of $\delta$. Therefore, $\delta$ is partitioned such that $\delta'$ computes two-thirds of the output channels and $\delta''$ computes one-third of the output channels. The $W$ parameters required for the operation are transferred to Processor 2, and $\delta''$ is computed in Processor 2. Thus we observe that \cref{fig:op_parallel}(d) has a perfectly balanced pipeline, leading to an increase in throughput and reduction in latency.

\begin{algorithm}[!tb]
{\scriptsize
    {\bf Input:} DFG of DNN, \#processors ($N_p$), processors \\
    {\bf Output:} Critical loops $C_{l}$,  processor allocation $P_a$
    \begin{algorithmic}[1]
        \STATE \textit{//Step 1: Find all $C_{loops}$ and DNN layers $L$ in the DFG}
        \label{alg:step1}
        \STATE $C_{l}$, $L$ = find\_critical\_loops(DFG)
        \STATE{\textit{//Step 2 starts here}}
        \STATE{\textit{//Profile: For each DNN layer $l$ in DFG find layer compute time $t_c$, fixed computation time $t_{fix}$, and flexible computation time $t_{flex}$. Store in $T_c$, $T_{fix}$, and $T_{flex}$}}
        \label{alg:step2}
        \STATE $T_c$, $T_{fix}$, $T_{flex}$ = profile(DFG)
        \STATE $T_{tot}$ = {\bf for} {$i$ in $T_c[i]$}  {\bf do} sum(T[i])
        \label{alg:ttot}
        \STATE{\textit{//Find maximum processor time $T_p$}}
        \STATE $T_p$  = $\frac{T_{tot}}{N_p}$ \\
        \label{alg:tp}
        \STATE{\textit{//Step 3 starts here}}
        \label{alg:step3}
        \STATE{\textit{//Initialize flag, processor index $p_{idx}$ to 0 and processor idle time $T_{idle}$ to $T_p$}}
        \STATE{flag = 0}
        \WHILE{flag $=$ 0}
            \STATE{$p_{idx}$ = 0; $T_{idle}$ = $T_p$}
            \FOR {each $l$ in reversed($L$)}
                \IF{$T_{flex}[l]$ $<$ $T_{idle}$}
                    \STATE{allocate $T_{flex}[l]$ to  processors[$p_{idx}$] and update $P_a$}
                    \label{alg:tf_a}
                    \STATE{$T_{idle} = T_{idle} - T_{flex}[l]$ }
                \ELSE
                    \STATE \textit{// Partition  $T_{flex}[l]$ ($\delta$) to $\delta'$ and $\delta''$ with operational parallelism (OP)}
                    \STATE {$\delta', \delta'' =$ OP$(T_{idle},T_{flex}[l])$}
                    \STATE{allocate $\delta''$ to processors[$p_{idx}$] and update $P_a$}
                    \STATE{$p_{idx}$ $=$ $p_{idx}$ $+ 1$}
                    \STATE{allocate $\delta'$ to processors[$p_{idx}$] and update $P_a$}
                    \STATE{$T_{idle} = T_p - \delta'$}
                \ENDIF
                \IF{$T_{fix}[l]$ $>$ $T_{idle}$}
                \label{alg:t_fix1}
                    \STATE{$p_{idx}$ $=$ $p_{idx}$ $+ 1$}
                \ENDIF
                \STATE{allocate $T_{fix}[l]$ to processors[$p_{idx}$] and update $P_a$}
                \STATE{$T_{idle} = T_{idle} - T_{fix}[l]$ }
                \label{alg:t_fix2}
            \ENDFOR
            \IF{$p_{idx}$ $>$  $N_p$}
                \STATE{\textit{//Relax the max processor time constraint and flags stays 0}}
                \STATE{$T_p$ $=$ $\alpha \times T_p$}
                \label{alg:alpha}
            \ELSE
                \STATE {flag $=$ 1}
            \ENDIF
        \ENDWHILE
        \RETURN $L_c$, $P_a$
    \end{algorithmic}
    \caption{Partitioning algorithm for balanced pipeline generation.}
    \label{alg:flow}
}
\end{algorithm}

\subsection{Scheduling and partitioning algorithms}

The heart of the problem is designing an algorithm to process a data-flow graph (DFG) and generate a schedule in a processor-constrained environment. The algorithm is designed to augment or improve upon existing parallelization techniques like PipeDream and GPipe by providing a more theoretical basis for partitioning the network. The algorithm serves two purposes. First, coarse-grained partitioning based on layers and inter-layer pipelining maximize throughput and minimize communication overhead. Second, a fine-grained schedule based on precedence graph and critical path minimizes latency. 

The proposed layer partitioning scheme can be described by~\cref{alg:flow}. The proposed algorithm leverages the techniques in the MARS algorithm~\cite{Mars95} to schedule feedback loops. The first step in the process is to evaluate and find all layers ($L$) and all the critical loops in the DFG (line~\ref{alg:step1}). The critical loop, along with its path, can be found in O($de_d$) time using the minimum cycle mean (MCM) algorithm~\cite{iteboundIto95}, where $d$ is the number of delays in the DFG and $e_d$ are all edges between the delays. In the second step, we run a profiler over the network and calculate the computation times, communication overhead, and memory overhead of each layer in the network. We then classify computations as movable or immovable based on the constraints on communication and memory (line~\ref{alg:step2}). If a computation's communication overhead and overhead memory fall below a threshold, we classify that computation as movable; otherwise, it is immovable. We return time taken for all movable computations as $T_{flex}$ and immovable computations as $T_{fix}$. For simplicity, we assume only movement from layer $l$ to layer $l+1$ is allowed. 
Taking the total computation time of the network ($T_{tot}$), we can determine the target workload of each processor or the maximum processor time ($T_p$) for the required number of processors ($N_p$) in lines \ref{alg:ttot} and \ref{alg:tp} of \cref{alg:flow}. 

In the third step of~\cref{alg:flow} (line~\ref{alg:step3}), we iterate through all the layers of the network in reverse and try to allocate it to processors as follows. As the flexible portion, $T_{flex}[l]$, can only be allocated to layer $l+1$, this is allocated first. If the computation can be accommodated entirely in the current processor $p_{idx}$, i.e., $T_{flex}[l]<T_{idle}$ we allocate this computation to the current processor, processors[$p_{idx}$] (line~\ref{alg:tf_a}). If this is not possible, we use operational parallelism to partition $\delta$ ($T_{flex}[l]$) into  $\delta'$ and $\delta''$ such that $\delta''$ fits within the remaining workload available in the processor. We then move to the next processor and assign the remaining computation $\delta'$. We then try to allocate the fixed portion of the computation in a similar manner. If the computation can be accommodated entirely in the current processor $p_{idx}$, i.e., $T_{flex}[l]<T_{idle}$, we allocate this computation to the current processor with index $p_{idx}$. If this is not possible, we move to the next processor and assign the computation (lines~\ref{alg:t_fix1}--\ref{alg:t_fix2}).
This process is repeated until all layers of the network have been assigned. If more processors are needed than those available, we relax the target workload requirements of each processor by a factor of $\alpha$ (line~\ref{alg:alpha}) to allow for longer compute times and restart at step 3. This process is repeated until the number of processors assigned matches the number of processors targeted. 

\cref{fig:scheduling} summarizes the differences between a traditional pipeline parallel schedule like PipeDream and a balanced, fine-grained pipeline schedule like LayerPipe. In the first stage, we derived the allocations of each processor, including the suggested operational parallelism from~\cref{alg:flow}. Using the information from the critical loops, we derive a schedule that prioritizes computations along the critical loop. Note that the critical loops generally run through the $\delta $ computation (\cref{fig:sample_network}) and the $G$ computation does not appear in the critical loop except in the first layer. Furthermore, as the only dependence of $\delta^{l-1}$ in layer $l$ from layer $l+1$ is $\delta^{l}$ we need not wait for the $G$ computation from layer $l+1$ to complete before starting layer $l$. Therefore, we can derive a schedule that prioritizes the $\delta$ computations over $G$ computations allowing for fine-grained pipelining that reduces the overall latency of the system. When operational parallelism is active, the two partitions $\delta'$ and $\delta''$ are independent; therefore, these can be computed in parallel, further reducing the system latency.

\begin{table}[!tb]
\caption{Network architecture for sample 4 convolution layer design.}
\label{tab:sample_arch}
\resizebox{\linewidth}{!}{%
\begin{tabular}{c|c|c|c|c|c}
\hline
Layer &Filter size & Input channels & Output channels & Padding & Stride \\ \hline
1 & 5 & 3   & 32  & 2 & 2 \\
2 & 5 & 32  & 64  & 2 & 2 \\
3 & 3 & 64  & 128 & 1 & 2 \\
4 & 3 & 128 & 128 & 1 & 1 \\ \hline
\end{tabular}%
}
\end{table}

\begin{table}[tb]
\caption{Summary of computation times and communication overheads for the sample 4-layer network.}
\centering
\label{tab:sample_summary}
\resizebox{0.9\linewidth}{!}{%
\begin{tabular}{c|cccc}
\hline
Layer & 1 & 2 & 3 & 4 \\ \hline
 & \multicolumn{4}{c}{Computation time (cycles)} \\ \cline{2-5} 
FP & $1.20\times10^{6} $ & $5.02\times10^{6} $  & $1.81\times10^{6} $ & $3.63\times10^{6} $  \\
BP\_G & $2.16\times10^{6} $ & $5.63\times10^{6} $  & $2.11\times10^{6} $ & $3.92\times10^{6} $  \\
BP\_$\delta$ & $4.01\times10^{7} $ & $2.01\times10^{7} $  & $7.23\times10^{6} $ & $3.63\times10^{6} $  \\
Total & $4.35\times10^{7} $ & $3.07\times10^{7} $  & $1.12\times10^{7} $ & $1.12\times10^{7} $  \\ \hline
 & \multicolumn{4}{c}{Communication overhead} \\ \cline{2-5} 
FP Overhead & 12.69MB & 6.57MB & 3.29MB & 3.52MB \\
BP Overhead & 4.59MB & 12.25MB & 6.13MB & 3.06MB \\
Additional Overhead & 0.07KB & 0.78KB & 0.56KB & 1.13KB \\ \hline
\end{tabular}%
}
\vspace{-1.0em}
\end{table}

\begin{table}[tbh]
\caption{Comparison of computation times in cycles of pipeline parallelism algorithms for the sample four-layer network.}
\label{tab:sample_dist}
\centering
\resizebox{0.8\linewidth}{!}{%
\begin{tabular}{l|ccc}
\hline
Processors & 1 & 2 & 3 \\ \hline
PipeDream & $4.35\times10^{7}$ & $3.07\times10^{7}$ & $2.23\times10^{7}$ \\ \hline
 \multicolumn{4}{c}{LayerPipe} \\ \hline
Assigned computation & $3.22\times10^{7}$ & $2.09\times10^{7}$ & $2.23\times10^{7}$ \\
Borrowed computation & $0.00\times10^{0}$ & $1.13\times10^{7}$ & $9.86\times10^{6}$ \\ \hline
Total & $3.22\times10^{7}$ & $3.22\times10^{7}$ & $3.22\times10^{7}$ \\ \hline
\end{tabular}%
}
\vspace{-1em}
\end{table}

 \begin{figure}[h]
    \centering
    \includegraphics[width=0.85\linewidth]{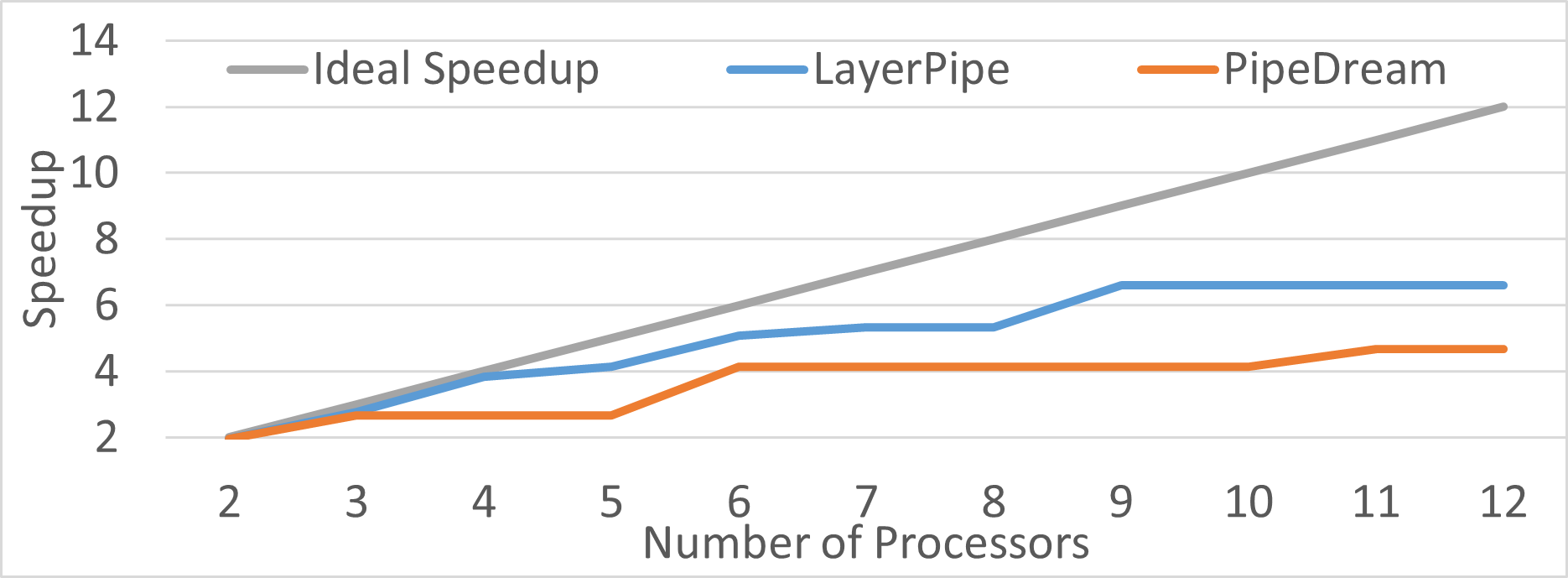}
    \caption{Comparison of speedups between LayerPipe and PipeDream with different number of processors on the convolutional layers of VGG16.}
    \label{fig:VGGl6_speedup}
\end{figure}

\section{Experimental evaluation}
\label{sec:results}
\subsection{Methodology}
In order to test the effectiveness of the proposed system, we benchmark the performance of LayerPipe against the standard pipeline parallelism algorithm, PipeDream~\cite{Pipedream}. The system simulated consists of multiple Processors that can communicate intermediate results among themselves without any external memory. The algorithm is evaluated by varying the number of processors while balancing the pipelines. Each processor consists of a single systolic array within a TPU or neural processing unit inside a GPU. For this experiment, it is assumed the systolic array is operated in a weight-stationary data-flow~\cite{SzeTutotrial}; however, the same techniques are applicable for other data-flows. In order to model the systolic array, we developed a python simulator based on the SCALESim~\cite{scalesim2019} library to estimate the computation times and communication requirements of the systolic array. The new simulator was validated against SCALESim, a cycle-accurate systolic array simulator verified against RTL simulations.  

To account for communication overheads, the simulator keeps track of two kinds of overheads. First, mandatory overheads, such as communication of activations and $\delta$ between processors, are required irrespective of the algorithm for any pipeline parallel design. Second, the additional overheads introduced by the LayerPipe algorithm are calculated. These overheads account for additional inputs that processors must communicate to support the LayerPipe algorithm, such as layer weights. 

The simulator tests whether the proposed algorithm is hardware agnostic by varying the systolic array size and batch size. The results are then averaged to obtain the final network performance results. 

\begin{figure*}
    \centering
    \begin{multicols}{2}

    \subfloat[a)]{\includegraphics[width=0.78\linewidth]{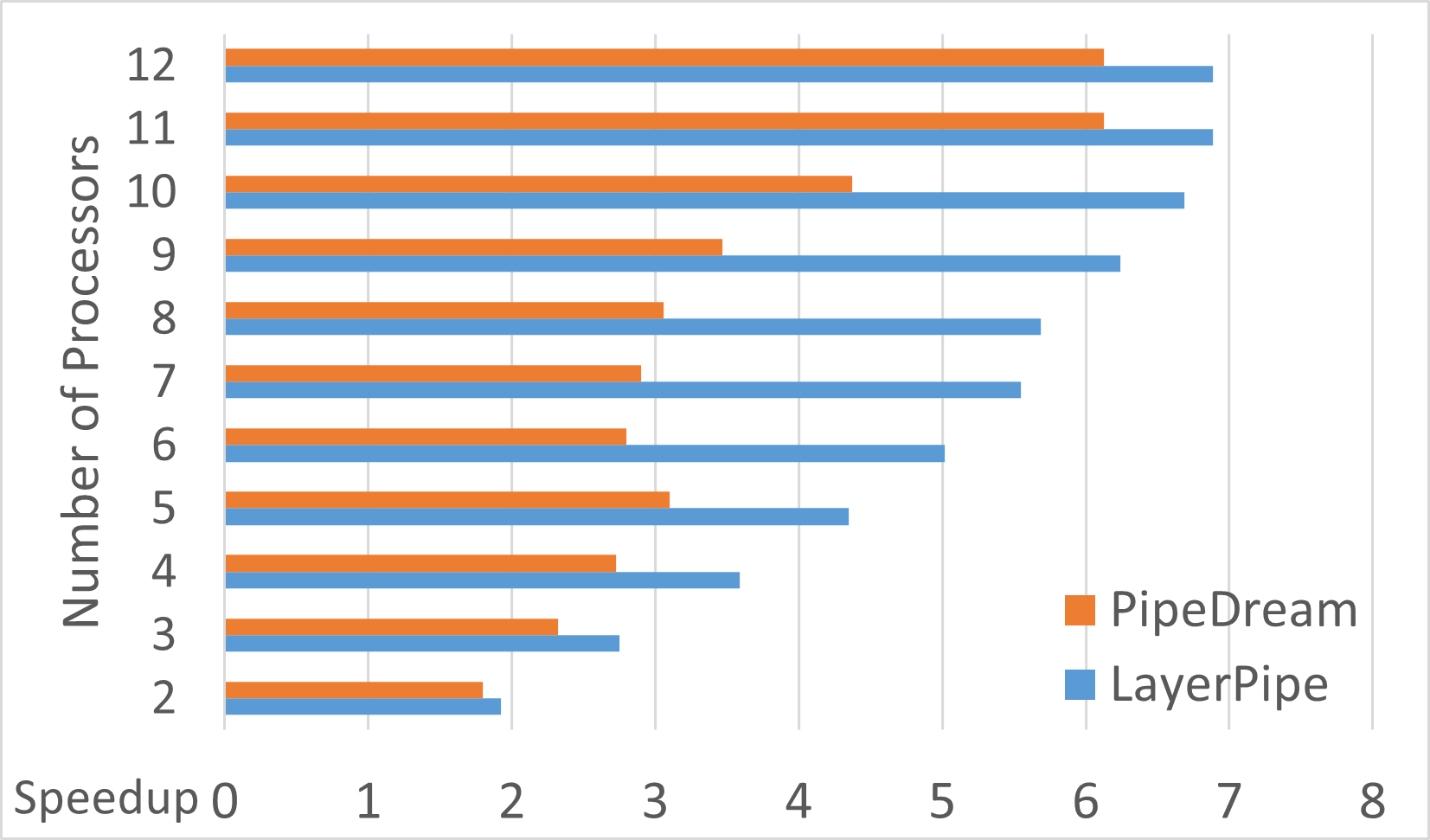}}\\
    \subfloat[c)]{\includegraphics[width=0.8\linewidth]{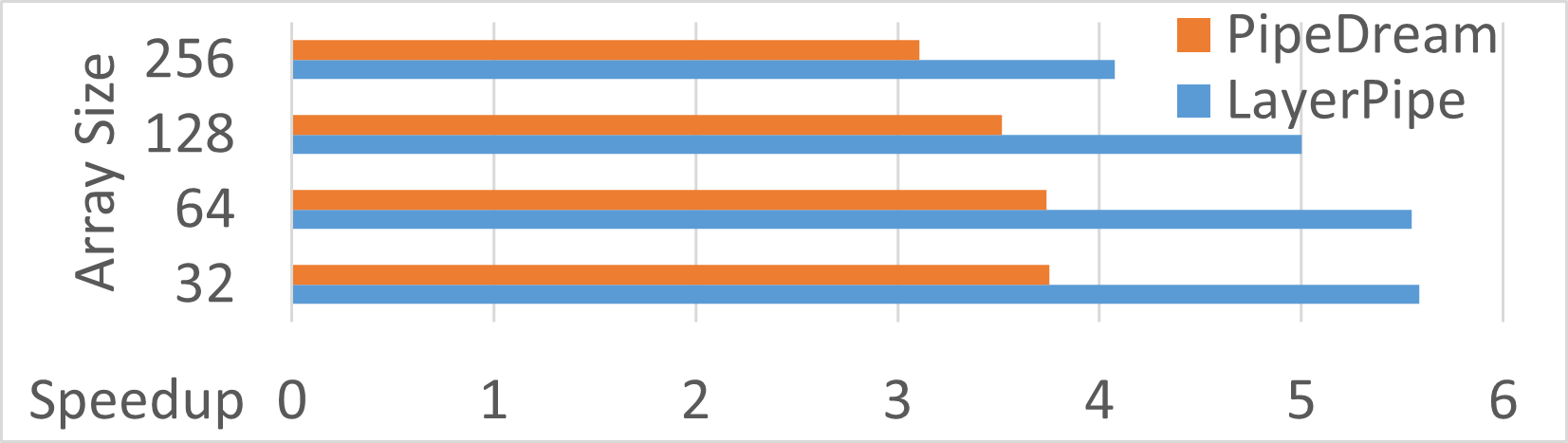}}
    
    \newpage
    \subfloat[b)]{\includegraphics[width=0.78\linewidth]{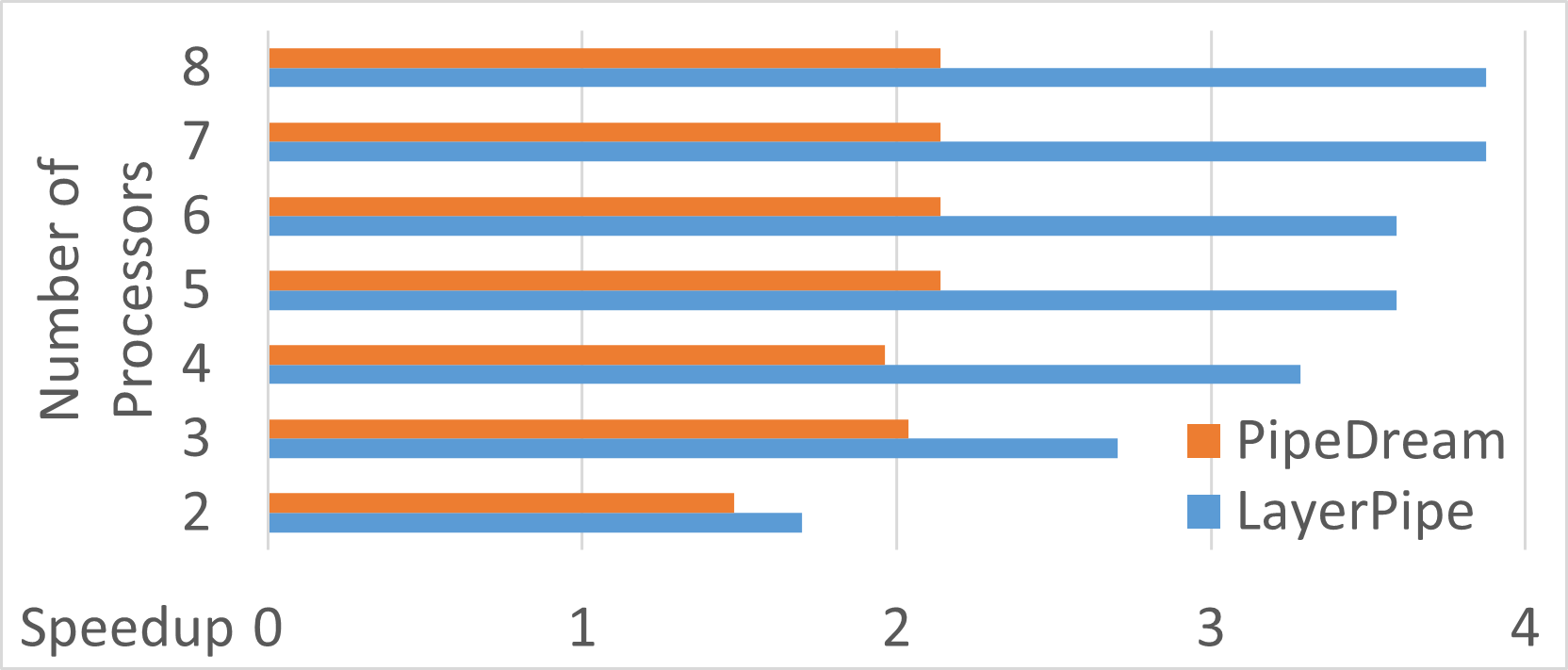}}\\    \vspace*{\fill}
    \subfloat[d)]{\includegraphics[width=0.78\linewidth]{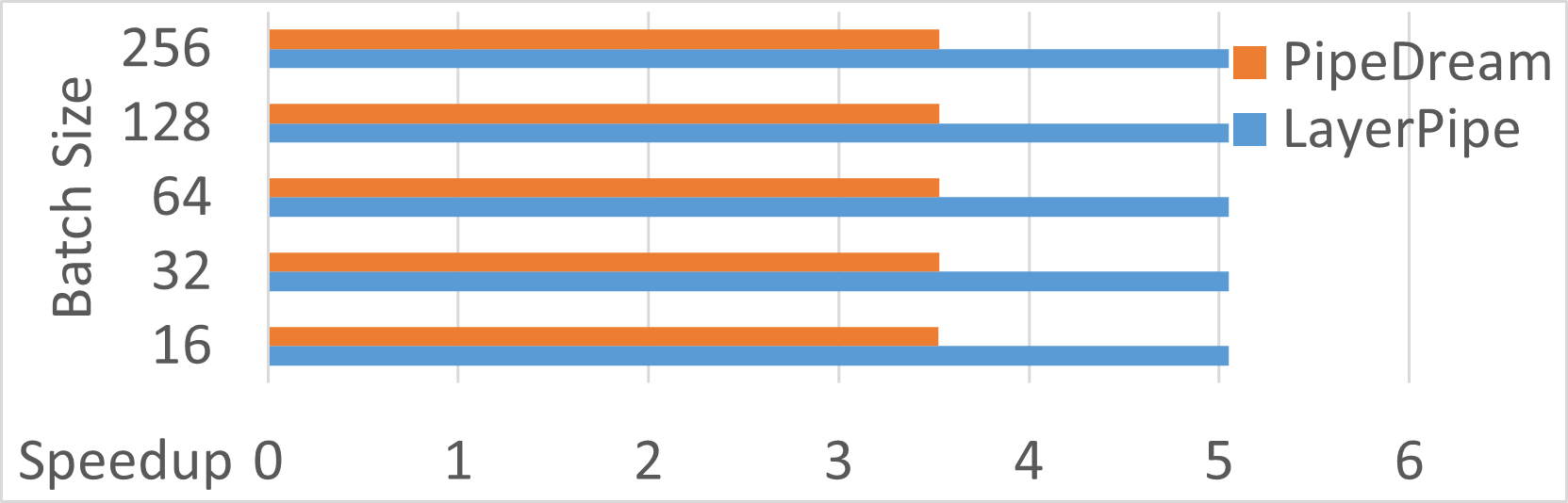}}
    
    \end{multicols}
    \caption{Performance comparison between LayerPipe and PipeDream for different number of processors on the convolutional layers of a) VGG16 b) ResNet50. The results were averaged over various batch sizes, systolic array sizes, and normalized to a single processor's performance. The individual distributions for VGG16 is shown for c) array size and d) batch size.}
    \label{fig:average_speedup}
    \vspace{-0.2em}
\end{figure*}

\begin{table}[h]
\caption{Comparison between the theoretical speedup of LayerPipe versus PipeDream for the convolution layer of VGG16. The results are averaged across systolic array and batch sizes.}
\label{tab:vgg_average}
\centering
\resizebox{0.9\linewidth}{!}{%
\begin{tabular}{c|ccc|c}
\hline
Processor & LayerPipe & PipeDream & Improvement & Communication Overhead \\ \hline
2 & 1.93 & 1.80 & 7.2\% & 0KB \\
3 & 2.75 & 2.32 & 18.4\% & 0KB\\
4 & 3.59 & 2.73 & 31.7\% & 0KB\\
5 & 4.35 & 3.10 & 40.2\% & 0.26KB\\
6 & 5.01 & 3.10 & 61.8\% & 0.01KB\\
7 & 5.55 & 3.10 & 79.0\% & 0.01KB\\
8 & 5.69 & 3.10 & 83.5\% & 0.01KB\\
9 & 6.24 & 3.47 & 80.0\% & 2.73KB\\
10 & 6.69 & 4.37 & 52.9\% & 2.61KB\\
11 & 6.88 & 6.13 & 12.4\% & 1.49KB\\
12 & 6.88 & 6.13 & 12.4\% & 0KB\\ \hline
\end{tabular}%
}
\vspace{-1em}
\end{table}

\subsection{Sample four-layer network}
The proposed algorithm is first evaluated on a simple network consisting of four convolutional layers of different sizes.  This provides a detailed demonstration of how the workload is scheduled between processors leading to a balanced schedule.
The parameters of the architecture of the sample network in~\cref{fig:sample_network} are described in~\cref{tab:sample_arch}.  Here the filter size, input channels, output channels, padding, and stride represent the parameters of the convolution layer. The input to the system is a $224\times224$ image with $3$ channels. The system consists of three systolic arrays, each with $32\time32$ processing elements, and the minibatch size is 32. The computation times for each of the layers and the communication overheads are listed in~\cref{tab:sample_summary}. FP refers to the forward pass, BP\_G and BP\_$\delta$ refer to the calculation of $G$ and $\delta$ in the backward pass. FP and BP overheads indicate how much additional communication is required if a pipeline stage is added immediately after and before the layer, respectively. These overheads are mandatory for any pipeline parallel design and are unavoidable. The additional overhead is the additional communication overhead required to support the LayerPipe algorithm. As shown in~\cref{tab:sample_summary}, the additional communication overhead of LayerPipe is insignificant.

\cref{tab:sample_dist} compares LayerPipe, the proposed operation scheduling, against PipeDream, a traditional pipeline parallel design for the sample network in a three-processor system. In PipeDream, layers 1 and 2 are mapped to their own processor, while layers 3 and 4 are assigned to the third processor. This assignment leads to severely imbalanced pipelines, as seen in the computation times of each processor in~\cref{tab:sample_dist}.  LayerPipe balances the workload across the processors by borrowing computation time from the previous layer.  In this sample network, the computation available to borrow is sufficient to balance the pipelines. This results in a $26\%$ improvement in the system throughput ($3.22\times10^{7}$ versus $4.35\times10^{7}$) with an additional communication overhead of $2.54$KB.

\subsection{Convolutional neural networks}

Extending the sample four-layer network analysis, we perform a detailed comparison between LayerPipe and PipeDream for VGG16 and ResNet50. The tests were performed by sweeping the systolic array size from $32\times32$ to $256\times256$ and the minibatch size from $16$ to $256$ in powers of 2. 
\cref{tab:vgg_average} compares the speedup and communication overhead between LayerPipe and PipeDream. LayerPipe achieves on average 43\% improvement over PipeDream with a maximum increase of $2.73$KB in the communication overhead. Additionally, it achieves greater than $80\%$ improvement with $8$ processors. Note, after $11$ processors, the ideal time for a single processor is less than the fixed computation time that cannot be borrowed from a layer. Therefore, this level provides the maximum speedup achievable, and further pipelining is no longer advantageous in both PipeDream and LayerPipe, as shown by the saturated curve in~\cref{fig:VGGl6_speedup}.

\cref{fig:average_speedup} summarizes the performance of LayerPipe versus PipeDream by varying the batch and systolic array sizes and testing it on VGG16 and ResNet50. Figs. 8(a) and (b) show the speedup comparison between LayerPipe and Pipedream by varying the number of processors from 2 to 12 while averaging the results across batch and array sizes. LayerPipe consistently outperforms PipeDream on all benchmarks. \cref{fig:average_speedup}(c) summarizes the performance of the two methods across different systolic array sizes for the processors. It is seen that using processors with smaller arrays improves the performance of both PipeDream and LayerPipe, but LayerPipe performance improvement is far more significant. A similar analysis in \cref{fig:average_speedup}(d) for batch size shows that the performance remains constant for the range of batch sizes tested. This indicates that the pipeline is very susceptible to the systolic array's size but independent of the batch size. 
\section{Conclusion}
\label{sec:conclusion}

This paper presented LayerPipe, a novel approach to achieve intra-layer and inter-layer optimization to generate balanced schedules for the pipelined design. 
LayerPipe achieves an average speedup of $25\%$ and upwards of $80\%$ with 7 to 9 processors with an insignificant communication overhead compared to PipeDream. The use of delayed gradient may lead to some performance degradation. Since the computations in PipeDream are functionally equivalent to the LayerPipe, any performance degradation in LayerPipe will be the same as in PipeDream. It has been shown that the use of {\em relaxed look-ahead} can overcome any degradation in delayed LMS~\cite{RLA}. Future work will design accelerators that incorporate relaxed look-ahead, can adapt to multi-GPU clusters, incorporate complex hierarchical communication models for overhead computations, and will address a detailed analysis of branches for complex DNN topologies.

\bibliographystyle{IEEEtran}
\balance
\bibliography{LayerPipe.bib}
\end{document}